\documentclass{ifacconf}
\usepackage{amsmath}     % For advanced math formatting
\usepackage{amssymb}     % For additional mathematical symbols
\usepackage{graphicx}    % For including graphics
\usepackage{multirow}    % For multirow tables
\usepackage[table,xcdraw]{xcolor} % For colored tables

\usepackage{natbib}      % For bibliography management
\usepackage{url}    % For hyperlinks, must be loaded last
\usepackage{float}       % For precise figure placement (e.g., [H] for forcing exact position)
\usepackage{subcaption}  % For subfigures and subcaptions
\begin{document}
\begin{frontmatter}

{\small\textit{This paper was accepted for publication by IFAC under a Creative Commons CC BY-NC-ND license and was presented at the IFAC Diabetes Technology Conference held on May 8–9, 2025, in Valencia, Spain.}}

\title{A Comparative Study of Transformer-Based Models for Multi-Horizon Blood Glucose Prediction \thanksref{footnoteinfo}} 
%: Exploring the Impact of History Length
\thanks[footnoteinfo]{Work funded by National Institute Health (NIH) grant 1R01DK133148.}

\author[First]{Meryem Altin Karagoz} 
\author[First]{Marc D. Breton}
\author[First]{Anas El Fathi} 

\address[First]{Center for Diabetes Technology, the University of Virginia, 
   Charlottesville, VA, USA, (e-mails: \{ssy4uh, mb6nt, fwt9vd\}@virginia.edu)}

\begin{abstract}               
Accurate blood glucose prediction can enable novel interventions for type 1 diabetes treatment including personalized insulin and dietary adjustments. Although recent advances in transformer-based architectures have demonstrated the power of attention mechanisms in complex multivariate time series prediction, their potential for blood glucose (BG) prediction remains underexplored. We present a comparative analysis of transformer models for multi-horizon BG prediction, examining forecasts up to 4 hours and input history up to 1 week. The publicly available DCLP3 dataset\footnote{\url{public.jaeb.org/datasets/diabetes}} (n=112) was split (80\%-10\%-10\%) for training, validation, and testing, and the OhioT1DM dataset \footnote{\url{smarthealth.cs.ohio.edu/OhioT1DM-dataset.html}} (n=12) served as an external test set. We trained networks with point-wise, patch-wise, series-wise and hybrid embeddings,  using CGM, insulin, and meal data. For short-term blood glucose prediction, Crossformer, a patch-wise transformer architecture, achieved a superior 30 minute prediction of RMSE (15.6 mg / dL on OhioT1DM). For longer-term predictions (1h, 2h, and 4h) PatchTST, another path-wise transformer, prevailed with the lowest RMSE (24.6 mg/dL, 36.1 mg/dL, and 46.5 mg/dL on OhioT1DM). In general, models that used tokenization through patches demonstrated improved accuracy with larger input sizes, with the best results obtained with a one-week history. These findings highlight the promise of transformer-based architectures for BG prediction by capturing and leveraging seasonal patterns in multivariate time-series data to improve accuracy.

\end{abstract}


\end{frontmatter}
%===============================================================================

\section{Introduction}
Individuals with type 1 diabetes (T1D) experience an autoimmune destruction of beta cells, making them unable to produce insulin. Insulin is an essential hormone for maintaining blood glucose (BG) levels within the healthy range of 70-140 mg/dL. Irregular BG levels can significantly affect quality of life in individuals with diabetes, leading to severe complications such as cardiovascular disease, kidney damage, nerve damage, and vision loss \citep{nathan1993effect}. To regulate BG levels and prevent hypo- (BG $<$ 70 mg/dL) and hyperglycemia (BG $>$ 180 mg/dL) events, individuals with T1D use lifelong insulin therapy via injections or pumps. Continuous glucose monitoring (CGM) systems have revolutionized T1D treatment by providing real-time glucose readings, e.g., every 5 minutes, and enabling advanced integrated systems such as automated insulin delivery systems.

CGM systems have significantly contributed to the advancement of BG prediction research. Reliable BG forecasting helps anticipate glucose fluctuation, allowing timely insulin and dietary adjustments to reduce glycemic risks and improve time in range (TIR, BG 70-180 mg/dL) \citep{jacobs2023artificial}. Physiological models for BG prediction use complex calculations to mimic metabolic systems, while data-driven methods identify patterns from CGM data without detailed physiological information. AI-based data-driven methods, such as autoregressive models, neural networks enhance accuracy without physiological parameters and effectively manage large datasets \citep{shuvo2023deep}.

Continuous Glucose Monitoring (CGM) provides detailed time-series data on glucose measurements, which enhances the ability of deep learning models to accurately predict future glucose levels by providing a large dataset for training. Deep neural networks, including Recurrent Neural Networks (RNNs), Convolutional Neural Networks (CNNs), and Transformers, have contributed significantly to the advancement of BG prediction techniques. \cite{zecchin2014jump} proposed a jump neural network for 30-minute BG predictions using historical CGM and carbohydrate data. \cite{zhu2020dilated} developed RNN models for short-term predictions by utilizing long-term dependencies. \cite{martinsson2020blood} developed an RNN model for prediction tasks of 60 minutes using only CGM data. LSTMs are favored to capture non-linear patterns and temporal dependencies, considering meals and insulin \citep{rabby2021stacked, mosquera2022incorporating}. Models often use multivariate inputs like carbohydrate intake, insulin dosage, and physical activity to enhance prediction accuracy. Hybrid models that integrate CNN and LSTM methodologies are designed to effectively capture both spatial and temporal dependencies \citep{jaloli2023long}. Furthermore, transformers are increasingly being utilized to manage complex dependencies and interactions \citep{FATHI2024245}, which makes them particularly well suited for long-term BG prediction, as exemplified in \citep{acuna2023analyzing, hakim2024significant, zhu2024population, xue2024bgformer}.

Transformer models show promise for BG prediction, but still face challenges, particularly with glucose variability beyond one hour. Determining the optimal history length—especially when meals and insulin are involved—remains a key issue. While transformers can capture long-term dependencies, the required historical data for comprehending glucose behavior remains uncertain. This research investigates the impact of history length on transformer performance for short- and long-term BG predictions using multivariate input.
\section{Methodology}
This study conducts a comparative analysis that employs Zero Order Hold as a baseline, along with DLinear and transformer-based models including Transformer, Crossformer, PatchTST, iTransformer, and TimeXer, over historical lengths HL = [4, 12, 24 hours, 1 week] and prediction horizons PH = [30, 60, 120, 240 minutes]. The performance of the models has been evaluated using the public datasets DCLP-3 and OhioT1DM. This research evaluates transformer models and explores history lengths to improve multi-horizon BG prediction accuracy and reliability, supporting effective diabetes management.

\begin{figure*}
    \centering
    \includegraphics[width=0.8\textwidth]{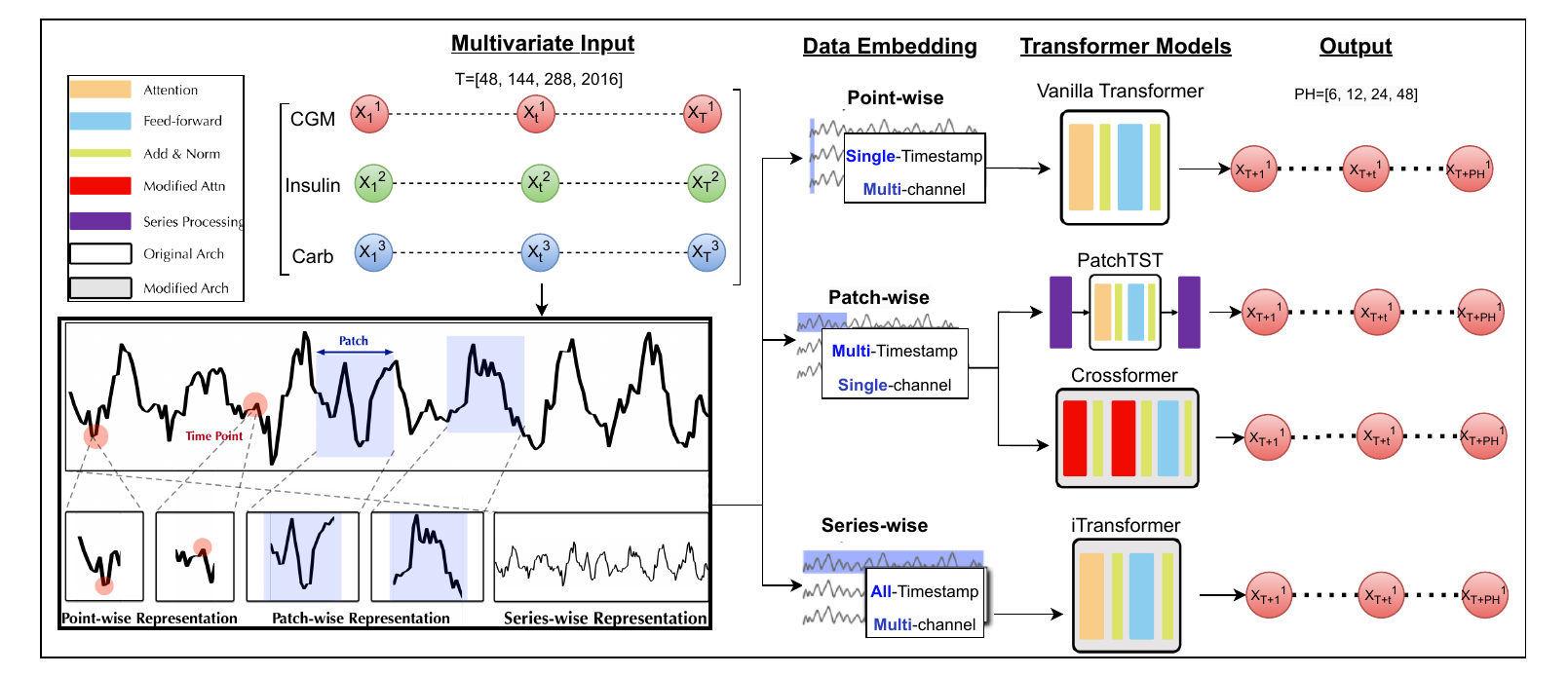}
    \caption{Illustration of the transformer models architecture for blood glucose prediction.}
    \label{fig:modelarch}
\end{figure*}

\subsection{Transformers and time-series forecasting}
Time series forecasting aims to explore patterns and relationships within historical temporal data to predict future behavior. The objective is to predict future values \([x_{T+1}, x_{T+2}, \dots, x_{T+PH}] \) given a time series \( X = \{x_1, x_2, \dots, x_T\} \in \mathbb{R}^{T \times C} \), where \( T \), \( C \) and \( PH \) represent the number of historical observations, the number of variables in the input and the prediction horizon, respectively. Transformer models became popular in time series analysis as in natural language processing and computer vision owing to their attention mechanism, calculated by Query-Key-Value (Q, K, V) as follows:

\begin{equation}
 {\scriptstyle \text{Attention}(\mathbf{Q}, \mathbf{K}, \mathbf{V})= \text{Softmax}\left( \frac{Q K^\top}{\sqrt{d_k}} \right) V }
\end{equation}

Current transformer-based time series models can be categorized according to their data representation (embedding) mechanisms: point-wise, patch-wise, and series-wise methods \citep{wang2024deep}. Therefore, this research focuses on cutting-edge transformer-based models, specifically Vanilla Transformer for point-wise, Crossformer-PatchTST for patch-wise, iTransformer for series-wise, and TimeXer as a hybrid approach. Figure 1 illustrates the data embedding techniques and model architectures.

The Vanilla transformer, introduced by \cite{vaswani2017attention}, is a sequence-to-sequence model with stacked encoder and decoder blocks. Each encoder block combines multihead self-attention, position-wise feedforward layers, residual connections, and layer normalization. The Vanilla transformer utilizes multihead attention rather than just a single attention mechanism. Queries (\( {Q} \in \mathbb {R}^{N \times D_k} \)), keys (\( {K} \in \mathbb{R}^{M \times D_k} \)), and values (\({V} \in \mathbb{R}^{M \times D_v} \)) are projected from a \( D_m \) dimensional input using \( H \) number of attention heads, each with learned projection matrices \( {W}_Q, {W}_K, {W}_V \) where N and M represent the lengths of queries and keys (or values). This multi-head attention mechanism enables parallel focus on different parts of the input. Attention is calculated for each projected set of queries, keys, and values. Ultimately, the results from all attention heads are combined and converted back into a \( D_m \)-dimensional output, as specified by following:
\begin{equation}
    {\scriptstyle\text{MultiHeadAttn}(\mathbf{Q}, \mathbf{K}, \mathbf{V}) = \text{Concat}(\text{head}_1,\ldots,\text{head}_H) {W}^O}
\end{equation}
where
\begin{equation}
    {\scriptstyle \text{head}_i = \text{Attention}(\mathbf{Q}\mathbf{W}_i^Q, \mathbf{K}\mathbf{W}_i^K, \mathbf{V}\mathbf{W}_i^V)}. 
\end{equation}

PatchTST \citep{nie2022time} employs a patch-based approach for time series with a channel-independent design, considering each channel as a univariate time series, and while sharing the same Transformer backbone, it maintains independent forward processes for all series. Patching divides each univariate time series into overlapping or non-overlapping patches of length P with a stride S. The number of patches N is given by formula \( N = \left\lfloor \frac{L-P}{S} \right\rfloor + 2 \), with L as the length of the series. This approach reduces the complexity of the memory and computational attention map by a factor of S. By reducing the quantity of input tokens, this approach enhances predictive accuracy over longer periods, diminishes computational load, and augments fine-tuning efficacy in self-supervised pre-training tasks. While PatchTST focuses on single-series dependencies, Crossformer, introduced by \citep{zhang2023crossformer}, captures interdependencies between multiple variables and time points. Crossformer is designed for multivariate time series forecasting with three key components: Dimension-Segment-Wise (DSW) embedding, Two-Stage Attention (TSA) layer, and Hierarchical Encoder-Decoder (HED) architecture. The HED architecture improves the accuracy of the prediction using DSW to identify dependencies in historical data and TSA to manage cross-temporal and cross-dimensional dependencies. The model employs a two-stage attention layer comprising cross-time and cross-dimensional stages to capture dependencies in patch tokens. Initially, it uses multihead self-attention on an embedded vector \( H \in \mathbb{R}^{N \times C \times d_{\text{model}}} \) for temporal dependencies, such as \( Z_{\text{time}} = \text{MSA}_{\text{time}}(H, H, H) \). In the dimension-wise stage, a learnable vector array \( R \) combines cross-dimensional data from \( Z_{\text{time}} \), as \( B = \text{MSA}_1^{\text{dim}}(R, Z_{\text{time}}, Z_{\text{time}} \), and the second stage refines this information as \( Z_{\text{dim}} = \text{MSA}_2^{\text{dim}}(Z_{\text{time}}, B, B) \).

iTransformer \citep{liu2023itransformer} employs inverted dimensions, attention, and feedforward networks in a standard transformer framework to capture multivariate correlations, preserving global series representations. It restructures the transformer's dimensional approach by encoding time points into variate tokens, then applying attention to uncover multivariate relationships. The feed-forward network is applied to each variate token individually, enhancing the model's ability to capture non-linear patterns. Tokenizing entire time series can further expand the receptive field to detect cross-variable dependencies. iTransformer handles multivariate data via embedded variate tokens ($H^{(i)} = \text{VariateEmbed}(X^{(i)} : \mathbb{R}^T \rightarrow \mathbb{R}^{d_{\text{model}}}$), where each variable \( X^{(i)} \) is projected through a trainable linear layer. Similarly, TimeXer \citep{wang2024timexer} incorporates exogenous variables with two representation levels—patch-level for endogenous data and series-level for exogenous data—and introduces an endogenous global token to strengthen temporal and cross-variable interactions. This approach boosts model generalization across variables and supports different look-back windows, making it a robust framework for time-series forecasting. 

In summary, point-wise transformer methods predict specific time points by using data embeddings and attention mechanisms to capture temporal relationships. Because the conventional attention mechanism grows quadratically with the sequence length, many transformer variants focus on improving attention efficiency. Meanwhile, patch-wise transformer models (e.g., PatchTST, Crossformer) chunk time-series data into patches, extracting local semantic features, and reducing complexity by lowering the number of input tokens. Series-wise methods, as exemplified by iTransformer, tokenize entire series and use global embeddings to capture mutual correlations, extending the receptive field. TimeXer combines both patch-wise and series-wise strategies, enhancing the capture of local and global temporal dynamics in multivariate time series. Accordingly, we conducted comparative analyzes using Transformer, Crossformer, PatchTST, iTransformer, and TimeXer to evaluate the impact of their embedding and attention mechanisms on BG prediction with multivariate inputs, as well as the effect of history length on model performance.

\subsection{Baseline methods}
The nontransformer-based Zero Order Hold (ZOH) and DLinear \citep{zeng2022dlinear} have been used as baseline models because of their simplicity, which allows a thorough evaluation of more advanced models. ZOH assumes that a variable's value stays constant until updated, serving as a basic benchmark for minimal change scenarios.  DLinear is a lightweight and effective time series forecasting model that decomposes data into trends and seasonality using simple linear layers. This approach enhances pattern recognition in tasks that focus on trends and seasonal differences. In summary, employing these baseline methods allows for a clearer understanding of the efficacy of advanced models in addressing complex data patterns.

\subsection{Training procedure}

We employ multivariate inputs, including carbohydrate information for meals, bolus, and CGM values, to predict target blood glucose levels. The input length, referred as history length, is established at {48, 144, 288, 2016} data points with 5-minute intervals, corresponding to durations of 4, 12, 24 hours, and 1 week, respectively. In addition, the prediction horizon is set at lengths of {6, 12, 24, 48} time points, equal to 30, 60, 120, and 240 minutes. Input and output sequences have been generated using a sliding-window methodology. The OhioT1DM dataset has been exclusively used for testing purposes. In this work, the complete OhioT1DM dataset was used for testing to evaluate the model's capacity for generalization on different treatment modalities after training on DCLP-3. The code is available in the GitHub account of the UVA Center for Diabetes and Technology\footnote{\footnotesize \url{https://github.com/Center-for-Diabetes-Technology/TransformerBasedBGPrediction}.}

%\section{Experiments}
\subsection{Datasets}
In this study, two data sets were used:
\begin{description}
    \item[DCLP3] This is a publicly available dataset courtesy of the JAEB Center for Helath Research\footnote{public.jaeb.org/datasets/diabetes}. Data include (n=112) adolescents (14–19 years) and adults (20 years and older) with T1D who participated in a six-month randomized, multicenter trial \citep{Brown2019}. Participants used Tandem t: slim X2 with Control-IQ Technology and Dexcom G6 CGM. Data set DCLP3 that has been divided by participants (to avoid training-testing leakage) into training, validation, and testing sets as ratio in (80-10-10\%) at subject level.
    \item[OhioT1DM] This is a publicly available dataset \footnote{smarthealth.cs.ohio.edu/OhioT1DM-dataset.html} with eight weeks of clinical trial for 12 patients with T1D (seven men and five women) aged between 20 and 80 years \citep{Marling2020}. Data were collected using Medtronic 530G insulin pumps and Medtronic Enlite CGM sensors, with additional daily events reported by patients through a smartphone app or fitness band. The entire data set was used for testing.
\end{description}
\subsection{Experimental Setup}
Models were developed using the PyTorch library and is executed on an A100-40GB GPU. Training is conducted for up to 10 epochs, incorporating early stopping to mitigate the risk of overfitting, with a batch size of 32 and a dropout rate of 0.1. The Adam optimizer was used across all models, with a learning rate of 1e-3. The MSE loss function is utilized as the default criterion. The ReduceLROnPlateau scheduler was implemented to adjust the learning rate in response to model performance, particularly when the validation loss plateaus.
\subsection{Performance Evaluation}
The evaluation of model performance has been conducted using widely recognized metrics: The root mean square error (RMSE), the mean absolute error (MAE), and the mean absolute relative difference (MARD). These metrics are computed for each subject using the predicted values ($y_{pred}$) \( \in \mathbb{R}^{N \times L} \) and actual ($y_{true}$) \( \in \mathbb{R}^{N \times L} \), where \( N \) is the number of samples and \(L \in \{6, 12, 24, 48\}\) is the prediction length, as demonstrated by equations 4,5,6, respectively.

\begin{equation}
 {\scriptstyle \text{RMSE} = \sqrt{\frac{1}{N}\sum_{i=1}^{N} \frac{1}{L} \sum_{j=1}^{L} \left( y_{\text{true},i,j} - y_{\text{pred},i,j} \right)^2}}
\end{equation}

\begin{equation}
{\scriptstyle \text{MAE} =\frac{1}{N} \sum_{i=1}^{N} \frac{1}{L} \sum_{j=1}^{L} \left| y_{\text{true},i,j} - y_{\text{pred},i,j} \right|}
\end{equation}

\begin{equation}
{\scriptstyle\text{MARD} = \frac{1}{N} \sum_{i=1}^{N} \frac{1}{L} \sum_{j=1}^{L} \frac{\left| y_{\text{true},i,j} - y_{\text{pred},i,j} \right|}{y_{\text{true},i,j}} \times 100}
\end{equation}

\section{Results and Discussion}
In this study, we conducted a comparative analysis with ZOH, point-wise methods (Dlinear, Transformer), patch-wise methods (PatchTST, Crossformer), series-wise methods (iTransformer), and hybrid methods (TimeXer) to evaluate their efficacy in both short-term and long-term predictions on the DCLP-3 and OhioT1DM datasets. The test results of these methods are reported in Table 1. Furthermore, Figure 2 provides a visualization of the RMSE results in all models with respect to the prediction horizons for varying historical lengths. The Crossformer model demonstrates superior performance in short-term predictions for 30-minute forecasts, achieving RMSE of 15.67, MAE of 10.13, MARD of 7.19 on the DCLP3 and RMSE of 15.81, MAE of 9.67, MARD of 7.02 on the OhioT1DM, utilizing 1 week of historical data. It is noteworthy that a CGM system with a MARD below 10\% is considered analytically effective. In the context of 60-minute predictions, the iTransformer model outperforms on the DCLP3 with RMSE of 24.97, MAE of 16.78, MARD of 12.03 while the PatchTST method achieves the lowest RMSE of 24.62, MAE of 16.38, MARD of 12.04 on the OhioT1DM, both with one week of historical data. In general, a one-week historical length emerges as the optimal compromise for transformer-based models when conducting 30- and 60-minute predictions.

\begin{table*}[ht]
\begin{center}
\caption{Multihorizon prediction results (RMSE, MAE and MARD) with mean $\pm$ std, on DCLP3 and OhioT1DM datasets using history length 4, 12, 24 hours, and 1 week. The minimum RMSE values for each PH across the various HLs are in red, while the second-lowest values are in green.}
%\fontsize{4.5pt}{4.5pt}\selectfont         % or \footnotesize, \scriptsize, etc.
%\setlength{\tabcolsep}{0.5pt}   % adjust column separation
%\renewcommand{\arraystretch}{1.2} % adjust row height if necessary
\resizebox{\textwidth}{!}{

\begin{tabular}{|c|c|c|ccc|ccc|ccc|ccc|}
\hline       
\multirow{2}{*}{} & \multirow{2}{*}{PH} & \multirow{2}{*}{Model} & \multicolumn{3}{c}{HL=4 hours} & \multicolumn{3}{|c|}{HL=12 hours} & \multicolumn{3}{c}{HL=24 hours} & \multicolumn{3}{|c|}{HL=1 week} \\
 &  &  & RMSE & MAE & MARD & RMSE & MAE & MARD & RMSE & MAE & MARD & RMSE & MAE & MARD \\ \hline
\multirow{28}{*}{\rotatebox{90}{DCLP3}} & \multirow{7}{*}{\rotatebox{90}{30 minutes}} & ZOH & 27.93 ± 3.60 & 20.07 ± 2.45 & 13.94 ± 2.02 & 27.93 ± 3.60 & 20.07 ± 2.45 & 13.93 ± 2.02 & 27.93 ± 3.60 & 20.07 ± 2.45 & 13.93 ± 2.02 & 27.93 ± 3.61 & 20.08 ± 2.45 & 13.94 ± 2.02 \\
 &  & DLinear & 16.43 ± 1.92 & 10.47 ± 1.07 & 7.33 ± 0.87 & 16.45 ± 1.90 & 10.59 ± 1.08 & 7.48 ± 0.86 & 16.44 ± 1.93 & 10.44 ± 1.10 & 7.24 ± 0.81 & 16.48 ± 1.91 & 10.65 ± 1.11 & 7.49 ± 0.80 \\
 &  & Transformer & 17.70 ± 1.87 & 12.11 ± 1.06 & 8.38 ± 0.86 & 17.78 ± 1.70 & 12.61 ± 0.97 & 8.65 ± 0.74 & 17.88 ± 1.82 & 12.58 ± 1.16 & 9.53 ± 1.40 & 19.46 ± 1.70 & 14.40 ± 1.08 & 9.49 ± 0.52 \\
 &  & {\color[HTML]{FF0000} Crossformer} & 16.16 ± 1.71 & 10.80 ± 1.00 & 7.81 ± 1.01 & 18.09 ± 1.57 & 13.62 ± 1.00 & 10.41 ± 1.30 & 16.95 ± 1.49 & 12.22 ± 0.90 & 9.28 ± 1.12 & {\color[HTML]{FF0000} 15.67 ± 1.74} & 10.13 ± 1.02 & 7.19 ± 0.84 \\
 &  & PatchTST & 16.27 ± 1.86 & 10.26 ± 1.04 & 7.12 ± 0.82 & 16.88 ± 1.88 & 11.60 ± 1.21 & 8.32 ± 0.83 & 16.28 ± 1.81 & 10.85 ± 1.12 & 7.71 ± 0.79 & 16.46 ± 1.80 & 10.97 ± 1.07 & 7.48 ± 0.70 \\
 &  & {\color[HTML]{006400}iTransformer} & 16.47 ± 1.90 & 10.48 ± 1.08 & 7.28 ± 0.85 & 16.50 ± 1.91 & 10.73 ± 1.12 & 7.47 ± 0.84 & {\color[HTML]{006400}15.82 ± 1.79} & 10.14 ± 1.03 & 7.10 ± 0.77 & 15.96 ± 1.78 & 10.48 ± 1.05 & 7.42 ± 0.76 \\
 &  & TimeXer & 16.62 ± 1.82 & 11.00 ± 1.07 & 7.74 ± 0.76 & 16.45 ± 1.87 & 10.87 ± 1.11 & 7.58 ± 0.75 & 16.14 ± 1.81 & 10.63 ± 1.09 & 7.57 ± 0.75 & 16.75 ± 1.81 & 11.58 ± 1.19 & 8.34 ± 0.77 \\ \cline{2-15} 
 & \multirow{7}{*}{\rotatebox{90}{60 minutes}} & ZOH & 44.42 ± 5.49 & 32.90 ± 3.84 & 23.05 ± 3.25 & 44.42 ± 5.49 & 32.90 ± 3.85 & 23.05 ± 3.25 & 44.42 ± 5.49 & 32.90 ± 3.85 & 23.05 ± 3.25 & 44.43 ± 5.50 & 32.91 ± 3.85 & 23.05 ± 3.25 \\
 &  & DLinear & 26.35 ± 3.16 & 17.52 ± 1.87 & 12.45 ± 1.66 & 26.32 ± 3.17 & 17.54 ± 1.90 & 12.47 ± 1.62 & 26.22 ± 3.15 & 17.55 ± 1.92 & 12.54 ± 1.53 & 26.55 ± 3.17 & 17.70 ± 1.93 & 12.32 ± 1.43 \\
 &  & Transformer & 29.43 ± 3.48 & 21.52 ± 2.35 & 14.11 ± 1.30 & 27.83 ± 3.00 & 19.88 ± 1.85 & 13.34 ± 1.31 & 27.43 ± 2.99 & 19.79 ± 1.87 & 13.44 ± 1.37 & 29.16 ± 3.18 & 21.24 ± 2.04 & 13.88 ± 1.16 \\
 &  & Crossformer & 25.30 ± 2.90 & 16.93 ± 1.74 & 12.25 ± 1.82 & 25.83 ± 2.88 & 17.59 ± 1.75 & 12.84 ± 1.77 & 25.19 ± 2.90 & 16.94 ± 1.76 & 12.24 ± 1.66 & 26.17 ± 2.88 & 17.82 ± 1.66 & 12.07 ± 1.44 \\
 &  & PatchTST & 26.68 ± 3.14 & 17.46 ± 1.86 & 12.20 ± 1.51 & 26.13 ± 3.04 & 17.65 ± 1.92 & 12.52 ± 1.45 & 25.29 ± 2.91 & 17.07 ± 1.86 & 12.34 ± 1.45 & 25.29 ± 2.96 & 17.23 ± 1.93 & 12.42 ± 1.46 \\
 &  & {\color[HTML]{FF0000}iTransformer} & 26.24 ± 3.11 & 17.18 ± 1.86 & 12.04 ± 1.52 & 25.55 ± 3.03 & 16.84 ± 1.85 & 11.84 ± 1.42 & {\color[HTML]{006400}25.04 ± 2.91} & 16.57 ± 1.78 & 11.74 ± 1.37 & {\color[HTML]{FF0000}24.97 ± 2.88} & 16.78 ± 1.79 & 12.03 ± 1.34 \\
 &  & TimeXer & 26.60 ± 3.05 & 17.86 ± 1.85 & 12.59 ± 1.41 & 25.45 ± 2.94 & 17.23 ± 1.83 & 12.18 ± 1.32 & 25.36 ± 2.90 & 17.23 ± 1.84 & 12.26 ± 1.34 & 25.10 ± 2.91 & 17.23 ± 1.91 & 12.48 ± 1.45 \\  \cline{2-15} 

 & \multirow{7}{*}{\rotatebox{90}{120 minutes}} & ZOH & 62.36 ± 6.98 & 47.47 ± 5.06 & 33.53 ± 3.74 & 62.36 ± 6.99 & 47.46 ± 5.07 & 33.52 ± 3.75 & 62.36 ± 6.98 & 47.46 ± 5.07 & 33.52 ± 3.75 & 62.37 ± 6.98 & 47.47 ± 5.06 & 33.53± 3.74 \\
 &  & DLinear & 37.90 ± 4.14 & 26.83 ± 2.65 & 19.55 ± 2.60 & 37.76 ± 4.23 & 26.70 ± 2.74 & 19.39 ± 2.47 & 37.58 ± 4.24 & 26.61 ± 2.82 & 19.38 ± 2.25 & 37.50 ± 4.27 & 26.87 ± 2.95 & 19.79 ± 2.12 \\
 &  & Transformer & 37.02 ± 4.01 & 26.27 ± 2.59 & 19.08 ± 2.72 & 37.07 ± 4.08 & 26.24 ± 2.61 & 18.90 ± 2.64 & 38.70 ± 4.23 & 28.54 ± 2.71 & 19.72 ± 2.16 & 37.62 ± 3.78 & 26.88 ± 2.52 & 20.29 ± 2.98 \\
 &  & Crossformer & 36.63 ± 3.95 & 25.91 ± 2.53 & 19.11 ± 2.90 & 36.62 ± 3.93 & 26.21 ± 2.52 & 19.41 ± 2.71 & 36.01 ± 3.97 & 25.35 ± 2.58 & 18.59 ± 2.39 & 35.67 ± 4.02 & 25.14 ± 2.69 & 18.41 ± 2.28 \\
 &  & {\color[HTML]{FF0000}PatchTST} & 39.74 ± 4.32 & 27.41 ± 2.77 & 19.19 ± 2.12 & 37.34 ± 4.22 & 25.95 ± 2.82 & 18.45 ± 2.09 & 36.10 ± 4.07 & 25.15 ± 2.74 & 17.97 ± 2.08 & {\color[HTML]{FF0000}35.59 ± 4.07} & 25.27 ± 2.82 & 18.32 ± 2.05 \\
 &  & {\color[HTML]{006400}iTransformer} & 38.55 ± 4.18 & 26.57 ± 2.69 & 18.86 ± 2.12 & 36.51 ± 4.13 & 25.26 ± 2.72 & 17.93 ± 2.02 & {\color[HTML]{006400}35.65 ± 4.07} & 24.79 ± 2.72 & 17.75 ± 2.01 & 36.17 ± 4.10 & 25.64 ± 2.81 & 18.57 ± 1.96 \\
 &  & TimeXer & 39.09 ± 4.32 & 27.27 ± 2.81 & 19.38 ± 2.05 & 37.10 ± 4.22 & 26.10 ± 2.85 & 18.74 ± 2.04 & 35.91 ± 4.09 & 25.36 ± 2.80 & 18.23 ± 1.98 & 36.01 ± 4.13 & 25.22 ± 2.80 & 18.22 ± 1.97 \\ \cline{2-15} 
 & \multirow{7}{*}{\rotatebox{90}{240 minutes}} & ZOH & 74.45 ± 9.22 & 57.57 ± 7.04 & 40.75 ± 4.07 & 74.45 ± 9.22 & 57.57 ± 7.04 & 40.75 ± 4.07 & 74.46 ± 9.21 & 57.58 ± 7.03 & 40.75 ± 4.06 & 74.46 ± 9.19 & 57.58 ± 7.031 & 40.75 ± 4.03 \\
 &  & DLinear & 46.67 ± 5.51 & 34.34 ± 3.48 & 24.62 ± 3.05 & 46.59 ± 5.32 & 34.55 ± 3.55 & 25.13 ± 2.85 & 45.85 ± 5.46 & 33.94 ± 3.74 & 24.76 ± 2.62 & 45.49 ± 5.54 & 33.57 ± 3.97 & 24.12 ± 2.39 \\
 &  & Transformer & 45.40 ± 5.20 & 33.50 ± 3.44 & 24.58 ± 3.48 & 47.70 ± 5.83 & 34.97 ± 3.74 & 24.11 ± 2.77 & 46.22 ± 5.13 & 34.18 ± 3.37 & 25.05 ± 3.37 & 46.24 ± 5.40 & 34.07 ± 3.65 & 24.41 ± 2.95 \\
 &  & Crossformer & 45.62 ± 5.16 & 33.57 ± 3.29 & 24.64 ± 3.48 & 45.23 ± 5.21 & 33.38 ± 3.38 & 24.46 ± 3.02 & 44.62 ± 5.21 & 32.96 ± 3.54 & 24.28 ± 2.74 & 44.32 ± 5.36 & 32.90 ± 3.86 & 24.43 ± 2.69 \\
 &  & {\color[HTML]{FF0000}PatchTST} & 50.99 ± 5.75 & 36.96 ± 3.95 & 26.23 ± 2.44 & 47.16 ± 5.68 & 34.52 ± 4.03 & 24.56 ± 2.46 & 44.86 ± 5.46 & 32.77 ± 3.86 & 23.64 ± 2.46 & {\color[HTML]{FF0000}43.61 ± 5.30} & 31.83 ± 3.78 & 22.95 ± 2.38 \\
 &  & {\color[HTML]{006400}iTransformer} & 49.36 ± 5.49 & 35.62 ± 3.71 & 25.23 ± 2.27 & 46.20 ± 5.48 & 33.57 ± 3.85 & 23.96 ± 2.44 & 44.44 ± 5.38 & 32.26 ± 3.80 & 23.14 ± 2.40 & {\color[HTML]{006400}43.78 ± 5.34} & 31.97 ± 3.82 & 23.06 ± 2.33 \\
 &  & TimeXer & 49.86 ± 5.70 & 36.24 ± 3.94 & 25.86 ± 2.37 & 46.15 ± 5.54 & 33.51 ± 3.89 & 23.84 ± 2.36 & 44.45 ± 5.43 & 32.35 ± 3.84 & 23.11 ± 2.35 & 43.96 ± 5.42 & 32.53 ± 3.98 & 23.97 ± 2.56 \\ \hline
\multirow{28}{*}{\rotatebox{90}{OhioT1DM}} & \multirow{7}{*}{\rotatebox{90}{30 minutes}} & ZOH & 25.59 ± 3.90&	17.75 ± 2.56&	12.19 ± 1.97&	25.60 ± 3.90&	17.75 ± 2.56&	12.19 ± 1.96&	25.61 ± 3.87&	17.76 ± 2.54&	12.19 ± 1.96	&25.62 ± 3.69	&17.75 ± 2.42&	12.16 ± 1.86

\\
 &  & DLinear & 16.54 ± 2.82& 9.77 ± 1.30& 6.87 ± 0.95& 16.56 ± 2.79& 9.89 ± 1.28& 7.06 ± 0.97& 16.61 ± 2.82& 9.83 ± 1.32& 6.84 ± 0.91& 16.62 ± 2.81& 10.03 ± 1.33& 7.13 ± 0.94
\\
 &  & Transformer & 29.84 ± 6.77& 23.85 ± 6.56& 19.06 ± 5.44& 22.26 ± 4.27& 16.31 ± 4.16& 13.28 ± 3.86& 24.89 ± 4.16& 18.70 ± 3.68& 16.14 ± 4.28& 18.65 ± 2.13& 13.29 ± 1.15& 10.18 ± 1.71
\\
 &  & {\color[HTML]{FF0000}Crossformer} & 16.05 ± 2.51& 10.09 ± 1.24& 7.54 ± 1.26& 18.10 ± 2.31& 13.07 ± 1.27& 10.44 ± 1.78& 17.04 ± 2.28& 11.72 ± 1.16& 9.32 ± 1.51& {\color[HTML]{FF0000}15.81 ± 2.56} & 9.67 ± 1.28& 7.02 ± 1.14
\\
 &  & PatchTST & 15.96 ± 2.65& 9.30 ± 1.28& 6.43 ± 0.87& 17.20 ± 2.46& 11.45 ± 1.35& 8.37 ± 1.01& 16.35 ± 2.54& 10.38 ± 1.35& 7.47 ± 0.97& 16.70 ± 2.41& 10.80 ± 1.30& 7.44 ± 0.79
\\
 &  & {\color[HTML]{006400}iTransformer} & 16.32 ± 2.78& 9.62 ± 1.35& 6.68 ± 0.92& 17.06 ± 2.63& 10.80 ± 1.34& 7.62 ± 0.87& {\color[HTML]{006400}15.86 ± 2.59} & 9.52 ± 1.28& 6.71 ± 0.88& 20.20 ± 2.29& 13.56 ± 1.38& 10.52 ± 1.46
\\
 &  & TimeXer & 16.72 ± 2.60& 10.47 ± 1.33& 7.39 ± 0.96& 16.70 ± 2.50& 10.68 ± 1.30& 7.60 ± 0.83& 16.32 ± 2.56& 10.24 ± 1.33& 7.52 ± 0.89& 17.02 ± 2.41& 11.46 ± 1.34& 8.56 ± 1.02
\\ \cline{2-15} 
 & \multirow{7}{*}{\rotatebox{90}{60 minutes}} & ZOH & 40.89 ± 6.27	&29.67 ± 4.41	&20.61 ± 3.61	&40.89 ± 6.26	&29.67 ± 4.41	&20.61 ± 3.60	&40.90 ± 6.23	&29.68 ± 4.38	&20.61 ± 3.60	&40.93 ± 5.91	&29.67 ± 4.17	&20.57 ± 3.42
\\
 &  & DLinear & 25.63 ± 3.67& 16.60 ± 2.08& 11.94 ± 1.87& 25.65 ± 3.64& 16.66 ± 2.06& 12.02 ± 1.83& 25.63 ± 3.64& 16.74 ± 2.09& 12.18 ± 1.82& 26.13 ± 3.61& 17.09 ± 2.09& 12.01 ± 1.63
\\
 &  & Transformer & 33.59 ± 4.48& 26.32 ± 3.83& 21.03 ± 4.83& 33.35 ± 5.00& 25.58 ± 4.55& 21.61 ± 5.73& 31.14 ± 3.68& 23.50 ± 2.90& 18.29 ± 3.08& 29.83 ± 3.95& 21.85 ± 2.99& 16.97 ± 3.45
\\
 &  & {\color[HTML]{006400}Crossformer} & {\color[HTML]{006400}24.63 ± 3.38}& 16.27 ± 2.03& 12.24 ± 2.26& 25.05 ± 3.39& 16.82 ± 2.06& 12.91 ± 2.37& 24.73 ± 3.31& 16.54 ± 2.02& 12.40 ± 2.11& 26.32 ± 3.31& 18.18 ± 2.01& 12.51 ± 2.00
\\
 &  & {\color[HTML]{FF0000}PatchTST} & 25.45 ± 3.62& 16.09 ± 2.09& 11.27 ± 1.62& 25.72 ± 3.37& 17.14 ± 2.08& 12.25 ± 1.61& 24.84 ± 3.43& 16.42 ± 2.12& 12.03 ± 1.75& {\color[HTML]{FF0000}24.62 ± 3.51} & 16.38 ± 2.23& 12.04 ± 1.81
\\
 &  & iTransformer & 26.04 ± 3.86& 16.31 ± 2.20& 11.48 ± 1.68& 25.20 ± 3.38& 16.14 ± 1.97& 11.41 ± 1.43& 25.09 ± 3.47& 16.35 ± 2.07& 11.83 ± 1.61& 26.51 ± 3.11& 18.15 ± 1.97& 13.91 ± 1.80
\\
 &  & TimeXer & 26.23 ± 3.55& 17.10 ± 2.10& 11.97 ± 1.59& 25.76 ± 3.37& 17.33 ± 2.08& 12.34 ± 1.55& 25.42 ± 3.41& 17.08 ± 2.14& 12.28 ± 1.62& 24.68 ± 3.35& 16.62 ± 2.10& 12.15 ± 1.71
\\ \cline{2-15} 
 & \multirow{7}{*}{\rotatebox{90}{120 minutes}} & ZOH & 60.47 ± 9.14&	45.53 ± 6.86&	32.45 ± 5.92	&60.48 ± 9.13	&45.53 ± 6.85	&32.45 ± 5.90	&60.50 ± 9.08&	45.55 ± 6.82	&32.45 ± 5.89	&60.54 ± 8.59	&45.54 ± 6.46	&32.38 ± 5.59

\\
 &  & DLinear & 37.58 ± 4.71& 26.51 ± 3.06& 19.76 ± 3.36& 37.53 ± 4.68& 26.46 ± 3.06& 19.69 ± 3.20& 37.64 ± 4.63& 26.63 ± 3.07& 19.92 ± 3.07& 37.86 ± 4.60& 27.12 ± 3.09& 20.60 ± 2.93
\\
 &  & Transformer & 42.72 ± 4.66& 33.08 ± 4.01& 27.36 ± 5.92& 43.62 ± 5.85& 33.56 ± 5.02& 26.87 ± 5.87& 42.81 ± 4.94& 32.97 ± 3.85& 25.48 ± 4.95& 46.18 ± 4.90& 36.62 ± 4.74& 31.25 ± 6.81
\\
 &  & {\color[HTML]{006400}Crossformer} & 36.49 ± 4.59& 25.90 ± 3.11& 19.94 ± 3.88& 36.50 ± 4.45& 26.19 ± 2.99& 20.23 ± 3.66& 36.28 ± 4.56& 25.60 ± 3.12& 19.57 ± 3.48&{\color[HTML]{006400}36.12 ± 4.66}& 25.46 ± 3.28& 19.34 ± 3.44
\\
 &  & {\color[HTML]{FF0000}PatchTST} & 39.00 ± 5.00& 26.60 ± 3.27& 18.82 ± 2.78& 37.49 ± 4.50& 25.92 ± 3.01& 18.68 ± 2.39& 36.61 ± 4.60& 25.35 ± 3.14& 18.24 ± 2.66& {\color[HTML]{FF0000}36.10 ± 4.59}& 25.47 ± 3.25& 18.70 ± 2.87
\\
 &  & iTransformer & 39.65 ± 5.04& 26.71 ± 3.20& 19.20 ± 2.63& 38.35 ± 4.72& 26.29 ± 3.11& 19.31 ± 2.44& 36.79 ± 4.64& 25.32 ± 3.14& 18.40 ± 2.63& 37.17 ± 4.56& 26.43 ± 3.22& 19.86 ± 2.70
\\
 &  & TimeXer & 39.49 ± 5.00& 27.06 ± 3.24& 19.03 ± 2.59& 38.34 ± 4.56& 27.05 ± 3.13& 19.48 ± 2.43& 37.23 ± 4.58& 26.19 ± 3.14& 18.80 ± 2.59& 36.91 ± 4.48& 25.80 ± 3.02& 19.19 ± 2.60
\\ \cline{2-15} 
 & \multirow{7}{*}{\rotatebox{90}{240 minutes}} & ZOH & 77.09 ± 10.64&	59.66 ± 8.06&	43.64 ± 7.78	&77.11 ± 10.61	&59.66 ± 8.04&	43.64 ± 7.75&	77.13 ± 10.54	& 59.67 ± 7.99&	43.64 ± 7.71	&77.16 ± 10.04	&59.65 ± 7.58&	43.55 ± 7.33 \\

 &  & DLinear & 48.60 ± 6.00& 36.11 ± 4.08& 26.62 ± 4.34& 48.37 ± 5.67& 36.09 ± 4.00& 27.11 ± 4.32& 48.12 ± 5.42& 35.93 ± 3.91& 27.14 ± 4.18& 48.17 ± 5.37& 35.98 ± 3.95& 26.81 ± 3.76
\\
 &  & Transformer & 55.12 ± 5.98& 45.25 ± 5.46& 37.96 ± 8.23& 51.92 ± 5.44& 40.44 ± 4.19& 31.25 ± 5.65& 54.44 ± 5.72& 43.98 ± 5.06& 36.76 ± 7.76& 52.00 ± 6.05& 41.07 ± 4.89& 32.07 ± 6.63
\\
 &  & Crossformer & 47.55 ± 5.61& 35.42 ± 3.95& 27.06 ± 5.02& 47.48 ± 5.46& 35.41 ± 3.86& 27.02 ± 4.53& 47.33 ± 5.31& 35.30 ± 3.91& 26.91 ± 4.41& 47.66 ± 5.33& 35.67 ± 4.09& 27.43 ± 4.46
\\
 &  & {\color[HTML]{FF0000}PatchTST} & 52.88 ± 6.09& 38.26 ± 4.31& 27.81 ± 3.99& 49.96 ± 5.83& 36.85 ± 4.29& 26.79 ± 3.39& 48.01 ± 5.46& 35.26 ± 4.05& 25.78 ± 3.55& {\color[HTML]{FF0000}46.49 ± 5.32}& 34.06 ± 3.94& 25.01 ± 3.63
\\
 &  & iTransformer & 54.55 ± 6.38& 39.16 ± 4.40& 28.80 ± 3.88& 50.80 ± 6.01& 37.09 ± 4.39& 27.34 ± 3.46& 48.33 ± 5.58& 35.29 ± 4.14& 25.96 ± 3.53& 47.37 ± 5.32& 34.90 ± 3.96& 25.86 ± 3.45
\\
 &  & {\color[HTML]{006400}TimeXer} & 53.28 ± 6.29& 38.36 ± 4.41& 27.02 ± 3.63& 50.27 ± 5.89& 36.47 ± 4.26& 25.39 ± 3.04& 48.52 ± 5.52& 35.44 ± 4.01& 25.03 ± 3.27& {\color[HTML]{006400}47.09 ± 5.38} & 34.94 ± 3.99& 26.09 ± 3.78\\
\hline
\end{tabular}
}
\end{center}
\end{table*}

\begin{figure*}[ht]
    \centering
    % Left subfigure (DCLP3 Results)
    \begin{subfigure}[b]{0.49\textwidth}
        \centering
            \includegraphics[width=0.9\linewidth]{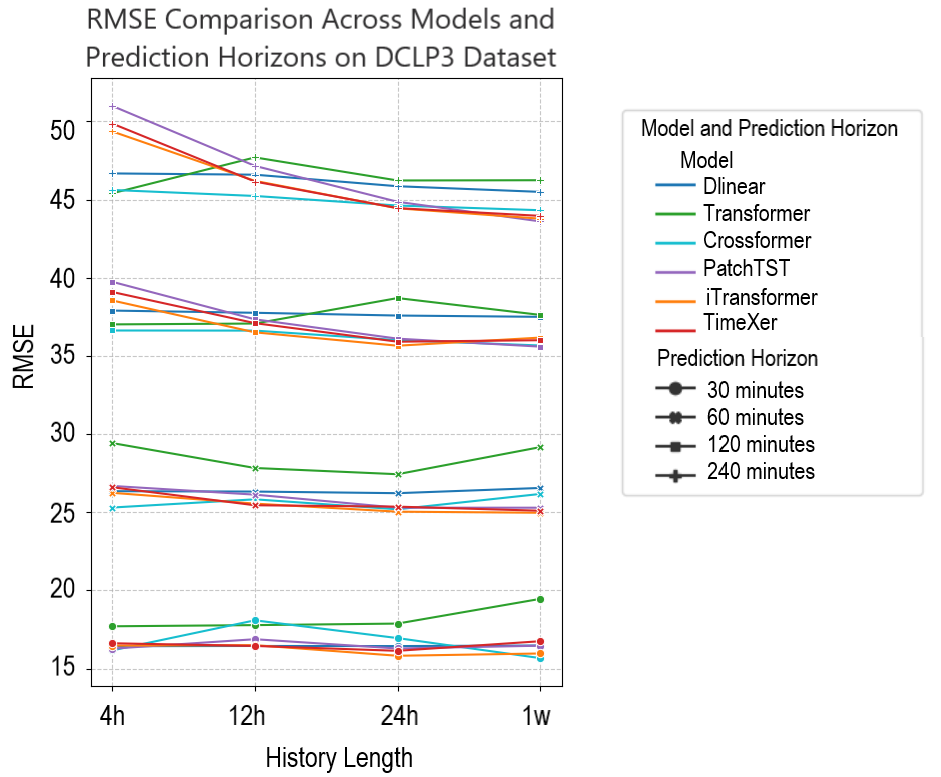}
            \caption{DCLP3 Results} % Ragged right for the left subfigure
        \label{fig:subfig1}
    \end{subfigure} %
    \hfill
    % Right subfigure (OhioT1DM Results)
    \begin{subfigure}[b]{0.49\textwidth}
        \centering
        \includegraphics[width=0.92\linewidth]{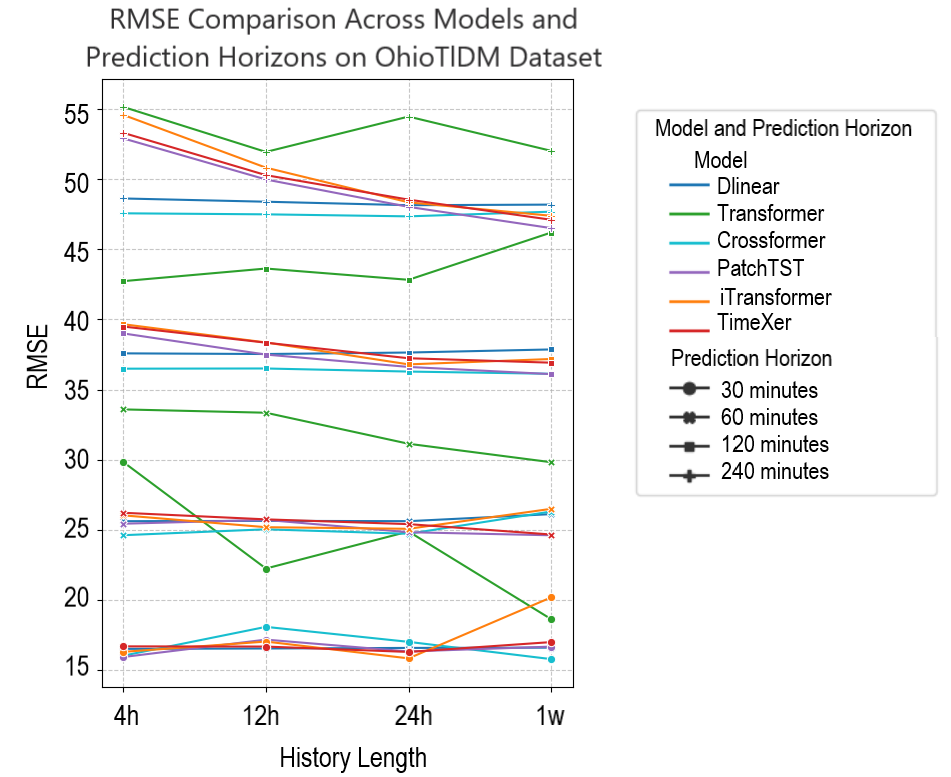}
        \caption{OhioT1DM Results} % Ragged right for the right subfigure
        \label{fig:subfig2}
    \end{subfigure}
    \caption{\raggedright Comparison of RMSE across models and prediction horizons for (a) OhioT1DM and (b) DCLP3 datasets. Each subfigure highlights the impact of history length and prediction horizon on model performance.} % Ragged right for the main figure caption
    \label{fig:combined_results}
\end{figure*}
In the context of long-term horizon prediction, the PatchTST model exhibits superior performance relative to other models for the 120-minute and 240-minute prediction horizon, when employing a history length of one week. Specifically, PatchTST achieves RMSE of 35.59, MAE of 25.27, MARD of 18.32 on the DCLP3 and RMSE of 36.10, MAE of 25.47, MARD of 18.70 on the OhioT1DM for the 120-minute prediction. For the 240-minute prediction, the model attains an RMSE of 43.61, MAE of 31.83, MARD of 22.95 on the DCLP3 and RMSE of 46.49, MAE of 34.06, MARD of 25.01 on the OhioT1DM. While patch-wise, series-wise, and hybrid transformers achieve remarkable improvement in long-horizon predictions with longer history, point-wise models (Transformer and DLinear) show limited improvement and are less sensitive to variations in history length (see Fig. 1). Furthermore, DLinear generally underperforms when compared to both patch-wise and series-wise transformer models, especially when considering a longer historical context. Whereas the history length increases, the performance of DLinear does not improve significantly, as DLinear lacks the capacity to capture complex temporal dependencies compared to advanced transformer-based models. However, the point-wise Transformer performs the worst on the OhioT1D dataset, indicating a lack of generalization ability from DCLP-3 to OhioT1DM. On the other hand, PatchTST, iTransformer, and TimeXer have more robust and competitive results with a longer history on both DCLP-3 and OhioT1DM datasets. In addition, Crossformer consistently shows a lower RMSE with a 4-hour history. As its history length increases, its performance improves further, making it increasingly competitive. The Crossformer appears an efficient model that consistently achieves a lower RMSE across different history lengths, even with a shorter history length for longer horizon prediction.
PatchTST, Crossformer, iTransformer, and TimeXer use novel attention mechanisms like patch-wise, two-stage, global-attention, and cross-attention, respectively, to capture multivariate relationships and long-term dependencies, improving performance in complex multivariate BG prediction. Future research should focus on mechanisms of attention to improve these models for efficiency and scalability, especially with exogenous variables or large datasets. ANOVA analysis was conducted to assess the differences among transformer models across prediction horizons and history lengths. High p-values (0.823 for 30 minutes, 0.596 for 60 minutes) did not indicate a significant variance in short-term forecasts. Conversely, low p-values (0.0327 for 120 minutes, 0.004 for 240 minutes) revealed significant differences in long-term performance, demonstrating transformers' effectiveness for long-term forecasting. As a result, while patch-wise and series-wise transformers outperform point-wise models, especially for long-term predictions by leveraging longer histories (with diminishing returns beyond 24 hours) and capturing complex temporal dependencies, point-wise models fall behind. 

\section{Conclusion}
This work examined how varying history lengths affect the performance of transformer-based methods, categorized by embedding strategies (point-wise, patch-wise, series-wise, and hybrid). A one-week history yields the most accurate short- and long-term forecasts overall. However, for patch-, series-, and hybrid-based transformers, using just 24 hours of history is both precise and efficient, with limited benefit from longer windows. Patching techniques enable these models to improve predictive accuracy by incorporating extended historical data without extra computational costs. In contrast, point-wise approaches underperform—especially over longer horizons—compared to patch- and series-wise methods. Although patch- and series-wise transformers gain accuracy from increased historical data, they still struggle with predictions beyond one-hour horizons, highlighting the need for new strategies to enhance long-term glucose forecasting. Further investigation into transformer-based methods is needed to enhance BG prediction, emphasizing model explainability and interpretability for therapeutic benefits.

\bibliography{ifacconf}

\begin{thebibliography}{23}
\providecommand{\natexlab}[1]{#1}
\providecommand{\url}[1]{\texttt{#1}}
\providecommand{\urlprefix}{URL }
\expandafter\ifx\csname urlstyle\endcsname\relax
  \providecommand{\doi}[1]{doi:\discretionary{}{}{}#1}\else
  \providecommand{\doi}{doi:\discretionary{}{}{}\begingroup \urlstyle{rm}\Url}\fi

\bibitem[{Acuna et~al.(2023)Acuna, Aparicio, and Palomino}]{acuna2023analyzing}
Acuna, E., Aparicio, R., and Palomino, V. (2023).
\newblock Analyzing the performance of transformers for the prediction of the blood glucose level considering imputation and smoothing.
\newblock \emph{Big Data and Cognitive Computing}, 7(1), 41.

\bibitem[{Brown et~al.(2019)Brown, Kovatchev, Raghinaru, Lum, Buckingham, Kudva, and et~al.}]{Brown2019}
Brown, S.A., Kovatchev, B.P., Raghinaru, D., Lum, J.W., Buckingham, B.A., Kudva, Y.C., and et~al. (2019).
\newblock Six-month randomized, multicenter trial of closed-loop control in type 1 diabetes.
\newblock \emph{New England Journal of Medicine}, 381(18), 1707--1717.
\newblock \doi{10.1056/NEJMoa1907863}.

\bibitem[{Fathi et~al.(2024)Fathi, Pryor, and Breton}]{FATHI2024245}
Fathi, A.E., Pryor, E., and Breton, M.D. (2024).
\newblock Attention networks for personalized mealtime insulin dosing in people with type 1 diabetes.
\newblock \emph{IFAC-PapersOnLine}, 58(24), 245--250.
\newblock \doi{https://doi.org/10.1016/j.ifacol.2024.11.044}.
\newblock 12th IFAC Symposium on Biological and Medical Systems BMS 2024.

\bibitem[{Hakim et~al.(2024)Hakim, Mahmud, Morsin, Setiawan, and Masdar}]{hakim2024significant}
Hakim, A.A., Mahmud, F., Morsin, M., Setiawan, R., and Masdar, A. (2024).
\newblock Significant of blood glucose time-series forecasting with temporal fusion transformer model for type 1 diabetes.
\newblock \emph{Journal of Advanced Research in Applied Sciences and Engineering Technology}, 312--324.

\bibitem[{Jacobs et~al.(2023)Jacobs, Herrero, Facchinetti, Vehi, Kovatchev, Breton, Cinar, Nikita, Doyle, Bondia et~al.}]{jacobs2023artificial}
Jacobs, P.G., Herrero, P., Facchinetti, A., Vehi, J., Kovatchev, B., Breton, M.D., Cinar, A., Nikita, K.S., Doyle, F.J., Bondia, J., et~al. (2023).
\newblock Artificial intelligence and machine learning for improving glycemic control in diabetes: Best practices, pitfalls, and opportunities.
\newblock \emph{IEEE reviews in biomedical engineering}, 17, 19--41.

\bibitem[{Jaloli and Cescon(2023)}]{jaloli2023long}
Jaloli, M. and Cescon, M. (2023).
\newblock Long-term prediction of blood glucose levels in type 1 diabetes using a cnn-lstm-based deep neural network.
\newblock \emph{Journal of diabetes science and technology}, 17(6), 1590--1601.

\bibitem[{Liu et~al.(2023)Liu, Hu, Zhang, Wu, Wang, Ma, and Long}]{liu2023itransformer}
Liu, Y., Hu, T., Zhang, H., Wu, H., Wang, S., Ma, L., and Long, M. (2023).
\newblock itransformer: Inverted transformers are effective for time series forecasting.
\newblock \emph{arXiv preprint arXiv:2310.06625}.

\bibitem[{Marling and Bunescu(2020)}]{Marling2020}
Marling, C. and Bunescu, R. (2020).
\newblock The ohiot1dm dataset for blood glucose level prediction: Update 2020.
\newblock In \emph{CEUR Workshop Proceedings}, volume 2675, 71. NIH Public Access.

\bibitem[{Martinsson et~al.(2020)Martinsson, Schliep, Eliasson, and Mogren}]{martinsson2020blood}
Martinsson, J., Schliep, A., Eliasson, B., and Mogren, O. (2020).
\newblock Blood glucose prediction with variance estimation using recurrent neural networks.
\newblock \emph{Journal of Healthcare Informatics Research}, 4, 1--18.

\bibitem[{Mosquera-Lopez and Jacobs(2022)}]{mosquera2022incorporating}
Mosquera-Lopez, C. and Jacobs, P.G. (2022).
\newblock Incorporating glucose variability into glucose forecasting accuracy assessment using the new glucose variability impact index and the prediction consistency index: An lstm case example.
\newblock \emph{Journal of Diabetes Science and Technology}, 16(1), 7--18.

\bibitem[{Nathan et~al.(1993)Nathan, Genuth, Lachin, Cleary, Crofford, Davis, Rand, Siebert et~al.}]{nathan1993effect}
Nathan, D., Genuth, S., Lachin, J., Cleary, P., Crofford, O., Davis, M., Rand, L., Siebert, C., et~al. (1993).
\newblock The effect of intensive treatment of diabetes on the development and progression of long-term complications in insulin-dependent diabetes mellitus.
\newblock \emph{The New England journal of medicine}, 329(14), 977--986.

\bibitem[{Nie et~al.(2022)Nie, Nguyen, Sinthong, and Kalagnanam}]{nie2022time}
Nie, Y., Nguyen, N.H., Sinthong, P., and Kalagnanam, J. (2022).
\newblock A time series is worth 64 words: Long-term forecasting with transformers.
\newblock \emph{arXiv preprint arXiv:2211.14730}.

\bibitem[{Rabby et~al.(2021)Rabby, Tu, Hossen, Lee, Maida, and Hei}]{rabby2021stacked}
Rabby, M.F., Tu, Y., Hossen, M.I., Lee, I., Maida, A.S., and Hei, X. (2021).
\newblock Stacked lstm based deep recurrent neural network with kalman smoothing for blood glucose prediction.
\newblock \emph{BMC Medical Informatics and Decision Making}, 21, 1--15.

\bibitem[{Shuvo and Islam(2023)}]{shuvo2023deep}
Shuvo, M.M.H. and Islam, S.K. (2023).
\newblock Deep multitask learning by stacked long short-term memory for predicting personalized blood glucose concentration.
\newblock \emph{IEEE Journal of Biomedical and Health Informatics}, 27(3), 1612--1623.

\bibitem[{Vaswani(2017)}]{vaswani2017attention}
Vaswani, A. (2017).
\newblock Attention is all you need.
\newblock \emph{Advances in Neural Information Processing Systems}.

\bibitem[{Wang et~al.(2024{\natexlab{a}})Wang, Wu, Dong, Liu, Long, and Wang}]{wang2024deep}
Wang, Y., Wu, H., Dong, J., Liu, Y., Long, M., and Wang, J. (2024{\natexlab{a}}).
\newblock Deep time series models: A comprehensive survey and benchmark.
\newblock \emph{arXiv preprint arXiv:2407.13278}.

\bibitem[{Wang et~al.(2024{\natexlab{b}})Wang, Wu, Dong, Qin, Zhang, Liu, Qiu, Wang, and Long}]{wang2024timexer}
Wang, Y., Wu, H., Dong, J., Qin, G., Zhang, H., Liu, Y., Qiu, Y., Wang, J., and Long, M. (2024{\natexlab{b}}).
\newblock Timexer: Empowering transformers for time series forecasting with exogenous variables.
\newblock \emph{arXiv preprint arXiv:2402.19072}.

\bibitem[{Xue et~al.(2024)Xue, Guan, and Jia}]{xue2024bgformer}
Xue, Y., Guan, S., and Jia, W. (2024).
\newblock Bgformer: An improved informer model to enhance blood glucose prediction.
\newblock \emph{Journal of Biomedical Informatics}, 157, 104715.

\bibitem[{Zecchin et~al.(2014)Zecchin, Facchinetti, Sparacino, and Cobelli}]{zecchin2014jump}
Zecchin, C., Facchinetti, A., Sparacino, G., and Cobelli, C. (2014).
\newblock Jump neural network for online short-time prediction of blood glucose from continuous monitoring sensors and meal information.
\newblock \emph{Computer methods and programs in biomedicine}, 113(1), 144--152.

\bibitem[{Zeng et~al.(2022)Zeng, Fu, Xu, Huang, Rong, Zhang, and Huang}]{zeng2022dlinear}
Zeng, A., Fu, Y., Xu, Q., Huang, Y., Rong, Y., Zhang, T., and Huang, J. (2022).
\newblock Are transformers effective for time series forecasting?
\newblock In \emph{Proceedings of the 39th International Conference on Machine Learning (ICML)}, 26731--26745.

\bibitem[{Zhang and Yan(2023)}]{zhang2023crossformer}
Zhang, Y. and Yan, J. (2023).
\newblock Crossformer: Transformer utilizing cross-dimension dependency for multivariate time series forecasting.
\newblock In \emph{The eleventh international conference on learning representations}.

\bibitem[{Zhu et~al.(2024)Zhu, Kuang, Piao, Zeng, Li, and Georgiou}]{zhu2024population}
Zhu, T., Kuang, L., Piao, C., Zeng, J., Li, K., and Georgiou, P. (2024).
\newblock Population-specific glucose prediction in diabetes care with transformer-based deep learning on the edge.
\newblock \emph{IEEE transactions on biomedical circuits and systems}, 18(2), 236--246.

\bibitem[{Zhu et~al.(2020)Zhu, Li, Chen, Herrero, and Georgiou}]{zhu2020dilated}
Zhu, T., Li, K., Chen, J., Herrero, P., and Georgiou, P. (2020).
\newblock Dilated recurrent neural networks for glucose forecasting in type 1 diabetes.
\newblock \emph{Journal of Healthcare Informatics Research}, 4, 308--324.

\end{thebibliography}

\end{document}